\documentclass[onecolumn,oneside,showpacs,%
  nofootinbib,aps,superscriptaddress,%
  eqsecnum,prd,notitlepage,showkeys,10pt]{revtex4-1}

\usepackage{amssymb}
\usepackage{amsmath}
\usepackage{graphicx}
\usepackage{dcolumn}
\usepackage{hyperref}
\usepackage[sc]{mathpazo} 
\usepackage[hmarginratio=1:1,top=32mm,columnsep=20pt,margin=1.8cm]{geometry} 

\begin{document}


\title{Monitoring temporal opacity fluctuations of large structures with muon radiography: a calibration experiment using a water tower}

\author{Kevin Jourde}
\affiliation{Institut de Physique du Globe de Paris (CNRS UMR 7154), Sorbonne Paris Cit\'{e}, Paris, France. }

\author{Dominique Gibert}
\affiliation{OSUR - G\'{e}osciences Rennes (CNRS UMR 6118), Universit\'{e} Rennes 1, Rennes, France.}
\affiliation{National Volcano Observatories Service, Institut de Physique du Globe de Paris (CNRS UMR 7154), Paris, France. }

\author{Jacques Marteau}
\affiliation{Institut de Physique Nucl\'eaire de Lyon, Univ Claude Bernard (UMR 5822 CNRS), Lyon, France.}

\author{Jean de Bremond d'Ars}
\affiliation{OSUR - G\'{e}osciences Rennes (CNRS UMR 6118), Universit\'{e} Rennes 1, Rennes, France.}

\author{Serge Gardien}
\affiliation{Institut de Physique Nucl\'eaire de Lyon, Univ Claude Bernard (UMR 5822 CNRS), Lyon, France.}

\author{Claude Girerd}
\affiliation{Institut de Physique Nucl\'eaire de Lyon, Univ Claude Bernard (UMR 5822 CNRS), Lyon, France.}

\author{Jean-Christophe Ianigro}
\affiliation{Institut de Physique Nucl\'eaire de Lyon, Univ Claude Bernard (UMR 5822 CNRS), Lyon, France.}

\keywords{Cosmic muons, Tomography, Monitoring}


\begin{abstract}
Usage of secondary cosmic muons to image the {geological structures density distribution} significantly developed during the past ten years. Recent applications demonstrate the {method interest} to monitor magma ascent and volcanic gas movements inside volcanoes. Muon radiography could be used  to monitor density variations in aquifers and the critical zone in the near surface. {However, the time resolution achievable by muon radiography monitoring remains poorly studied. It is biased by fluctuation sources exterior to the target, and statistically affected by the limited number of particles detected during the experiment. The present study documents these two issues within a simple and well constrained experimental context~: a water tower.}. {We use the} data to discuss the influence of atmospheric variability that perturbs the signal{, and} propose correction formulas to extract the {muon flux variations} related to {the water level changes}. Statistical developments establish the feasibility domain of muon radiography monitoring as a function of target thickness (i.e. opacity). Objects with a thickness comprised between $\approx 50 \pm 30 \; \mathrm{m}$ water equivalent correspond to the best time resolution. Thinner objects have a degraded time resolution that strongly depends on the zenith angle, whereas thicker objects (like volcanoes) time resolution does not.   
\end{abstract}

\maketitle


\section*{Introduction}

Using {the secondary cosmic rays muon component} to image geological bodies like volcano lava domes is the subject of increasing interest over the past ten years. Much like medical X-ray radiography, muon radiography aims at recovering the density distribution, $\rho$, inside the targets by measuring their screening effect on {the cosmic muons natural flux}. This approach was first tested by {George\cite{george1955cosmic} to measure the thickness of the geological overburden of a tunnel in Australia}, and later by Alvarez et al.\cite{alvarez1970search} who imaged the Egyptian Pyramid of Chephren to eventually find a hidden chamber. The method then stayed long dormant until recent years when, thanks to progress in electronics and particle detectors, field instruments were designed and constructed by several research teams worldwide \cite{tanaka2001development, marteau2014implementation}. Muon radiography experiments have successfully been performed on volcanoes where the hard muon component is able to cross several kilometres of rock \cite{nagamine1995geo, tanaka2001development, tanaka2005radiographic, lesparre2010geophysical, lesparre2012density, marteau2012muons, jourde2013experimental, portal2013inner, tanaka2014radiographic}. Applications to archaeology \cite{menichelli2007scintillating}, civil engineering (tunnels, dams) and environmental studies (near surface geophysics) are subject to active research, and monitoring of density changes in the near surface constitutes an important objective in hydrology and soil sciences.

The material property that can be recovered with muon radiography is the opacity, $\varrho$, which quantifies the amount of matter encountered by the muons along their travel path, $L$, across the volume to image,
\begin{equation}
\varrho = \int_L \rho(l) \times \mathrm{d}l.
\label{opacity1}
\end{equation}

Generally, the opacity is expressed in $[\mathrm{g.cm}^{-2}]$ or, equivalently, in centimetres water equivalent $[\mathrm{cm.w.e.}]$. Muons lose their energy through matter by ionisation processes \cite{nagamine2003introductory} at a typical rate of $2.5 \; \mathrm{MeV}$ per opacity increment of $1 \; \mathrm{g.cm}^{-2}$. They are relativistic leptons produced in the upper atmosphere at an altitude of about $16 \; \mathrm{km}$ \cite{gaisser1990cosmic, grieder2001cosmic}, and reach the ground after losing about $2.5 \; \mathrm{GeV}$ to cross the opacity of $10 \; \mathrm{m.w.e.}$ represented by the atmosphere. Muons travel along straight trajectories across low-density materials, including water, concrete and rocks, and scattering is significant only in high-density materials like lead and uranium \cite{nagamine2003introductory}. However, low-energy muons ($E \le 1 \; \mathrm{GeV}$) have strong scattering in almost all materials.

Muon radiography of kilometre-size objects like volcanoes involves the hard muonic component with energy above several hundredths of $\mathrm{GeV}$. In such cases, the {muons incident flux} may reasonably be considered stationary, azimuthally isotropic and to only depend on the zenith angle \cite{gaisser1990cosmic, lesparre2010geophysical}, and simple flux models can be used to determine the screening effects produced by the target to image\cite{tang2006muon, lesparre2010geophysical}. The situation is different in environmental and civil engineering applications where {the bodies have} low opacities {(i.e. several tens of $\mathrm{m.w.e.}$)} that can be crossed by the soft muonic component (i.e. several $\mathrm{GeV}$) which can no more be considered stationary and isotropic.

{The soft muon component main causes of non-stationarity and anisotropic characteristics\cite{gaisser1990cosmic, grieder2001cosmic} are pressure variations at the ground level \cite{yanchukovsky2007atmospheric}}, and geomagnetic storms together with solar coronal mass ejections (CME)\cite{kremer1999measurements, munakata2000precursors}. Both time-variations of atmospheric pressure and CME involve time constants of several hours or days \cite{chilingarian2005correlated} which are of a critical importance when monitoring fast density changes in low-opacity bodies. During CME and associated magnetic storms, one generally observes Forbush decreases that correspond to a deficit of cosmic rays of $1\%$ or $2\%$ at ground level, but variations up to $10\%$ have been {reported\cite{maghrabi2012thekacst}}. {The atmospheric pressure variations} typically produce {muon flux relative changes of} $0.1 \; \% . \mathrm{hPa}^{-1}$, i.e. {several percent variations} during perturbed meteorological conditions. Consequently, monitoring subtle density changes in low-opacity targets {necessitates a precise correction of the muon flux time-variations induced} by both atmospheric pressure variations and eventual intense geomagnetic events.

The present study aims at contributing to the {procedures development} to monitor density changes in low-opacity bodies. We apply and discuss a simple way to suppress atmospheric pressure effects from muon counting data. We present a controlled experiment performed on a water tank tower whose opacity fluctuates in a significant range ($3 \; \mathrm{m.w.e.} < \varrho < 5 \; \mathrm{m.w.e.}$) where atmospheric effects are expected to significantly perturb the {incident cosmic muons flux}. {We made measurements} during a {several weeks period} while the opacity remains steady at its maximum level before fluctuating. Meanwhile, water level in the tank, atmospheric pressure, and geomagnetic activity are monitored in order to evaluate their relative importance to produce {muon flux variations} across the water volume. Finally, a discussion {about} the time resolution in muon radiography monitoring is presented with a particular emphasis for {the low-opacity targets case}.

\section*{The SHADOW experiment}

\begin{figure}[ht] 
\centering
\includegraphics[width=0.6\linewidth]{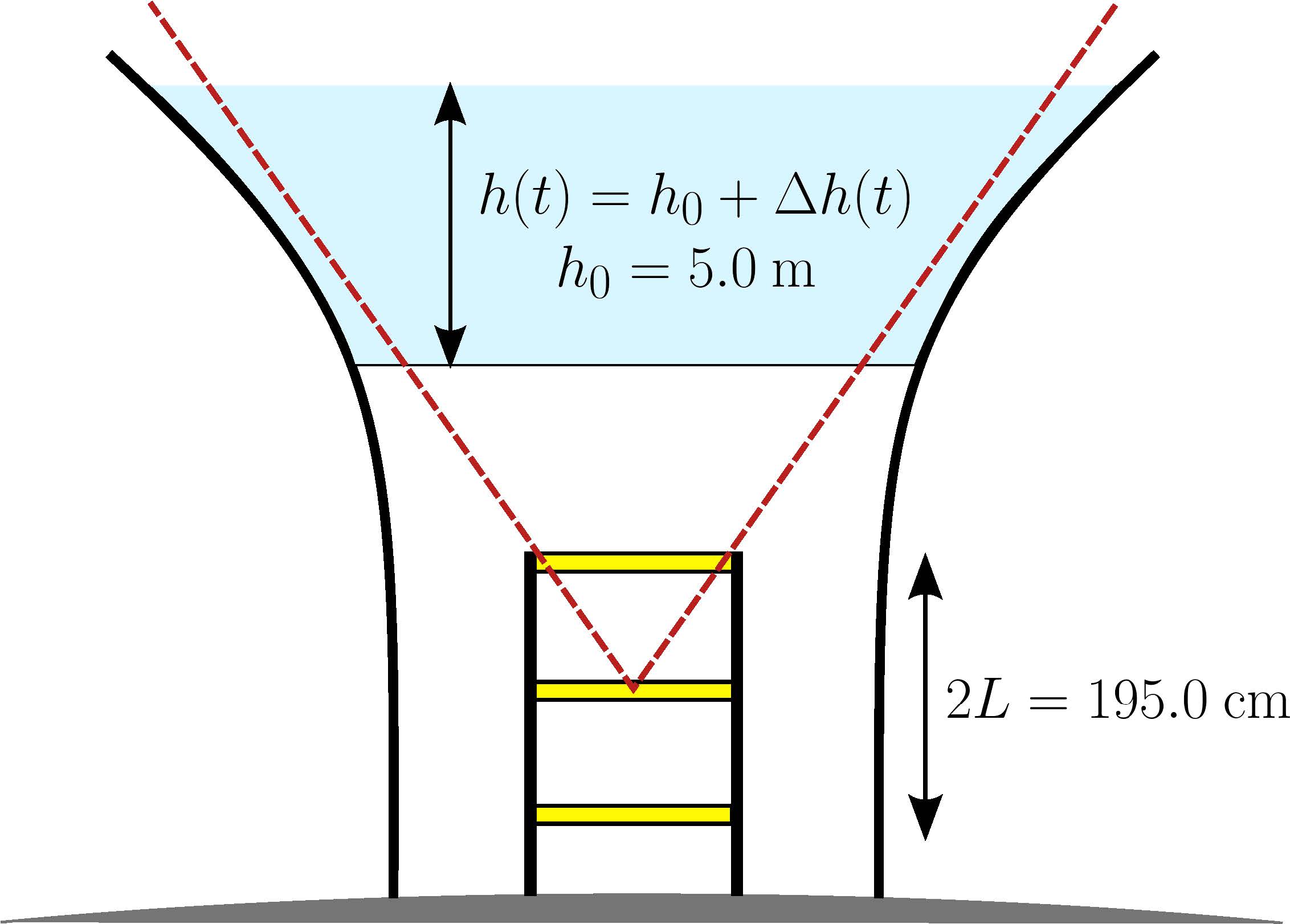}
\caption{Sketch of the SHADOW experiment. The three yellow rectangles are the detection matrices (each with $16 \times 16 \; \mathrm{pixels \, of} \; 5 \times 5 \; \mathrm{cm}^2$), the red dotted lines encompass the detection solid angle and the blue surface represents the water volume.}
\label{experimental_setup_figure}
\end{figure} 

The SHADOW experiment measured the {muonic component time-variations} while the water level varied in a water tower. For this purpose, {we placed} a muon telescope (its description is given in the Methods Section below) {along the tower symmetry axis} and below the tank. {We oriented} the instrument vertically (i.e. central zenith angle = 0) as shown in Figure~\ref{experimental_setup_figure} so that the apparent opacity is only {zenith angle (azimuthal invariance) and time (when the water level $h(t)$ is changing) dependent}. The water tower is located in Tignieu-Jameyzieu, France, a village located 20~kilometres East from Lyon (altitude $230 \; \mathrm{m}$ above sea level, $X_{UTM} = 31 \; 669490$, $Y_{UTM} = 5067355$). The distance between the upper and the lower matrices is set at $195 \hspace{1mm} \mathrm{cm}$ to cover a zenith angle range $0^\circ \le \theta \le 22.3^\circ$ such that all {the telescope $961$ lines} of sight pass through the water. The solid angle spanned by the telescope equals $\Omega_{\mathrm{int}} = 0.161 \hspace{1mm} \mathrm{sr}$, and the total effective acceptance $\mathcal{T}_{\mathrm{int}} = 630 \hspace{1mm} \mathrm{cm}^2\mathrm{sr}$.

The data acquisition started on November $21^{\mathrm{th}}$, $2014$ and stopped on January $22^{\mathrm{nd}}$, $2015$. While measuring the {muon flux under} the tank, the water level was monitored with {a several cm accuracy} every $5$ minutes by the company in charge of the tower (Syndicat Intercommunal des Eaux de Pont-de-Ch\'eruy -- SIEPC). These data are completed by {atmospheric pressure hourly measurements} at the nearby Saint Exupery airport {located $7.45 \; \mathrm{km}$ West of the water tower at an altitude of $248 \; \mathrm{m}$ above sea level.}

{During the measurement period, the geomagnetic activity was monitored using the Kp geomagnetic index published by the International Service of Geomagnetic Indices (ISGI) of the International Association of Geomagnetism and Aeronomy (IAGA). From November $21^{\mathrm{st}}$, $2014$ to January $22^{\mathrm{nd}}$, $2015$, a noticeable geomagnetic activity is reported for December $2014$ $7^{\mathrm{th}}, 12^{\mathrm{th}}, 22^{\mathrm{nd}}, 26^{\mathrm{th}}, 29^{\mathrm{th}}$ and January $2015$ $4^{\mathrm{th}}, 5^{\mathrm{th}}, 7^{\mathrm{th}}, 8^{\mathrm{th}}$ where the geomagnetic activity reached the "minor" G1 level\cite{poppe2000new} excepted in polar regions where the G3 level was observed. Sudden storm commencements (SSC) are reported on December $21^{\mathrm{st}}$ (19:11 UTC), $22^{\mathrm{nd}}$ (15:11 UTC), $23^{\mathrm{rd}}$ (11:15 UTC), and January $7^{\mathrm{th}}$ (06:14 UTC). It cannot be excluded that the geomagnetic activity at these dates produced small variations, at the fraction of percent level \cite{ishibashi2005observation}, of the muon flux measured during the SHADOW experiment. However, these variations are expected to occur only a few times in the data time-series and this sparsity prevents a detailed quantitative study to identify the corresponding signals.}

In the next section, we use hourly averages of these data series to document the relationship between the {muon flux time-variations} and those of both the atmospheric pressure and the water level in the tank.

\subsection*{Constant water level: {Atmospheric effects contribution}}

\begin{figure*}[bt] 
\centering
\includegraphics[width=0.8\linewidth]{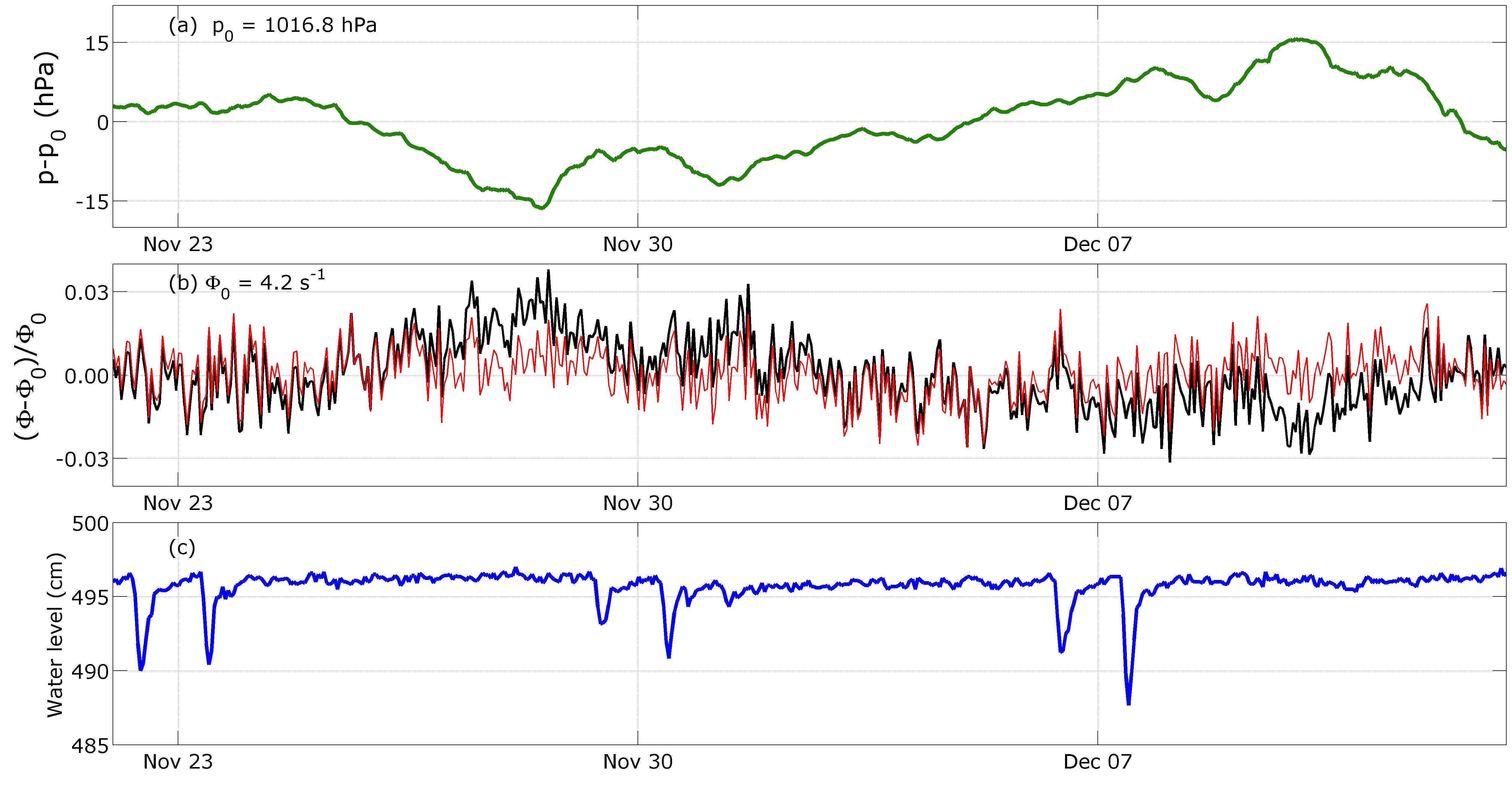}
\caption{
Hourly averages of the data acquired during the calibration period (November $22^{\mathrm{nd}}$ to December $13^{\mathrm{th}}$, $2014$): \textbf{(a)} {Atmospheric pressure time-variations relatively} to $p_0 = 1016.8 \; \mathrm{hPa}$. \textbf{(b)} {Muon flux relative raw time-variations} (black curve), {and corrected} from the atmospheric pressure influence (red curve). \textbf{(c)} {Water level variations}.
}
\label{dataCalibrationPeriod}
\end{figure*} 

\begin{figure}[bt] 
\centering
\includegraphics[width=0.55\linewidth]{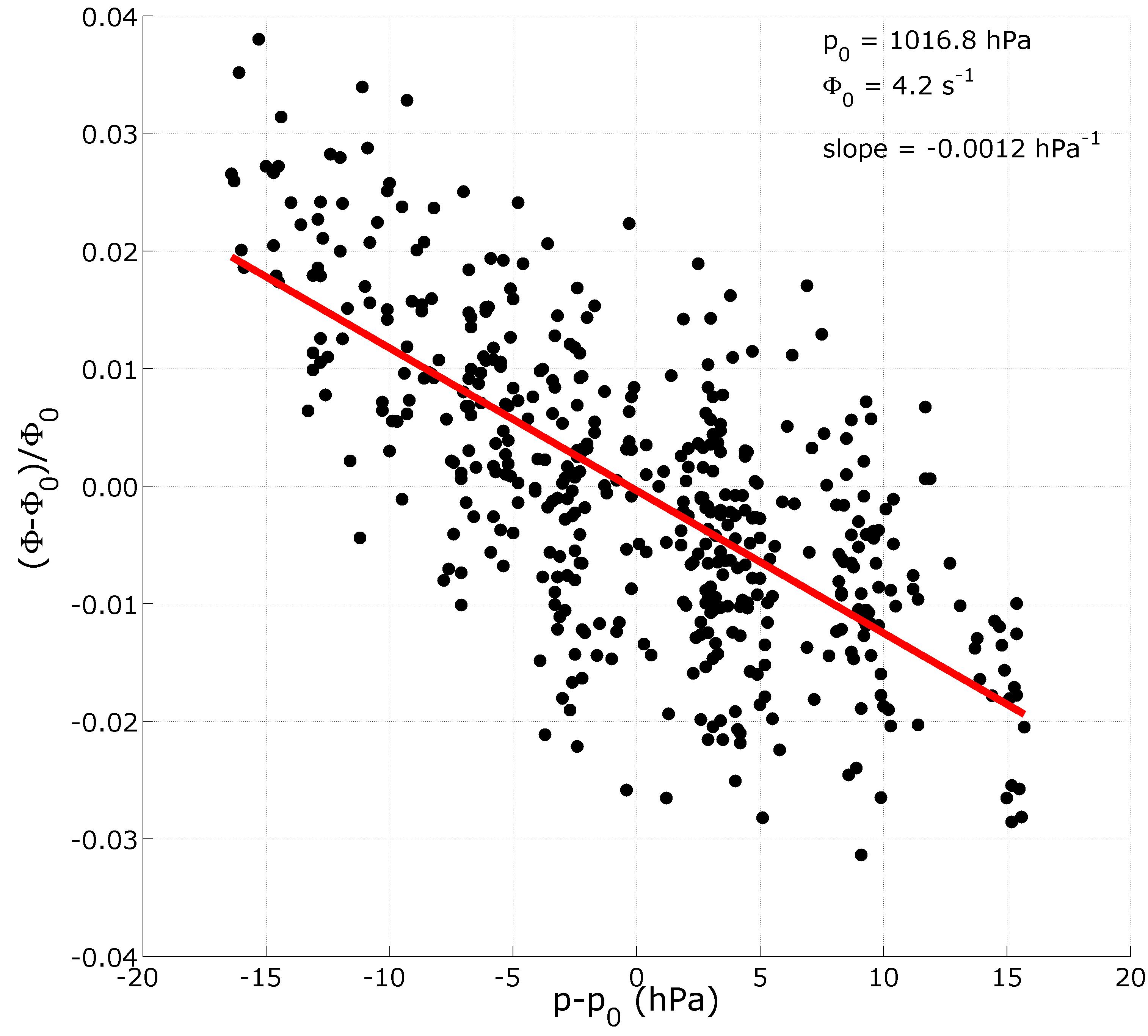}
\caption{
{Muon flux relative variations} versus atmospheric pressure deviation. The red line represents the best least-squares fit solution. Only the data points corresponding to a water level greater than $495 \; \mathrm{cm}$ have been kept to compute the line fit.
}
\label{fluxPressure}
\end{figure} 

We first consider the data acquired during the {measurement period first three weeks}, from November $22^{\mathrm{nd}}$ to December $13^{\mathrm{th}}$ $2014$, when the water level in the water tower remained almost constant at its maximum level $h_0 = 496 \hspace{1mm} \mathrm{cm}$ (Figure~\ref{dataCalibrationPeriod}c). Meanwhile, the atmospheric pressure varied by $\pm 15 \; \mathrm{hPa}$ with respect to a reference pressure $p_0 = 1016.8 \; \mathrm{hPa}$ (Figure~\ref{dataCalibrationPeriod}a). The {muon flux} shown on Figure (\ref{dataCalibrationPeriod}b) not only randomly fluctuates as expected for a Poissonian process but also contains long-period variations with an amplitude of less than $3 \%$. These long-period variations are clearly anti-correlated with those of the atmospheric pressure (Figure~\ref{dataCalibrationPeriod}a).

Since the water level is mainly constant during the considered period, {we expect} the {muon flux time variations} to be principally caused by atmospheric effects. The graph in Figure \ref{fluxPressure} represents the {muon flux hourly averages} with respect {to the atmospheric pressure from Saint Exupery airport}. In this graph, only the data points such that the water level $495 \hspace{1mm} \mathrm{cm} \le h \le 496 \hspace{1mm} \mathrm{cm}$ are retained. A least-squares fit to these points gives a negative slope $\beta_p = -0.0012 \; (0.0001) \; \mathrm{hPa}^{-1}$ where the value in parenthesis is the half-width of the $95\%$ confidence interval. {We performed the fit} by assigning to the relative flux averages a standard deviation $\sigma_\Phi = 0.0081$ derived from {the events arrival times statistics}. A standard deviation $\sigma_p = 1 \; \mathrm{hPa}$ is assigned to the atmospheric pressure data. The {linear fit residuals standard deviation}, $\sigma_r = 0.0093$, falls near $\sigma_\Phi$ and indicates that no higher-order fit is required. Consequently, in the remaining, we shall represent the atmospheric influence on the relative muon flux with a linear relationship,
\begin{equation}
\frac{\Phi - \Phi_0}{\Phi_0} = \beta_p \times (p - p_0).
\label{pressureCorrection}
\end{equation}

Dayananda \cite{dayananda2013understanding} uses the same kind of linear relation and finds $\beta_p = -0.0013 \; (0.0002) \; \mathrm{hPa}^{-1}$ from muon counts at the Earth's surface. Other authors \cite{sagisaka1986atmospheric, motoki2003precise, rigozo2013atmospheric} also find linear relationships with coefficients falling near $\beta_p = -0.001 \; \mathrm{hPa}^{-1}$. {We do not expect the barometric coefficient derived in the present study to be strictly equal to those obtained for other experiments since this coefficient is sensitive to the site location and especially to the telescope altitude\cite{dorman2004cosmic, chilingarian2011calculation}. However they should be in the same order of magnitude.}

The correction formula (\ref{pressureCorrection}) says that a $\Delta p = 10 \; \mathrm{hPa}$ increase of the atmospheric pressure induces a relative muon flux decreases of $1.2 \%$. Applying this correction to the muon flux data (black curve of Figure \ref{dataCalibrationPeriod}b) efficiently reduces the long-period variations (light red curve of Figure \ref{dataCalibrationPeriod}b).

\subsection*{Time-varying water level}

\begin{figure}[tb] 
\centering
\includegraphics[width=0.8\linewidth]{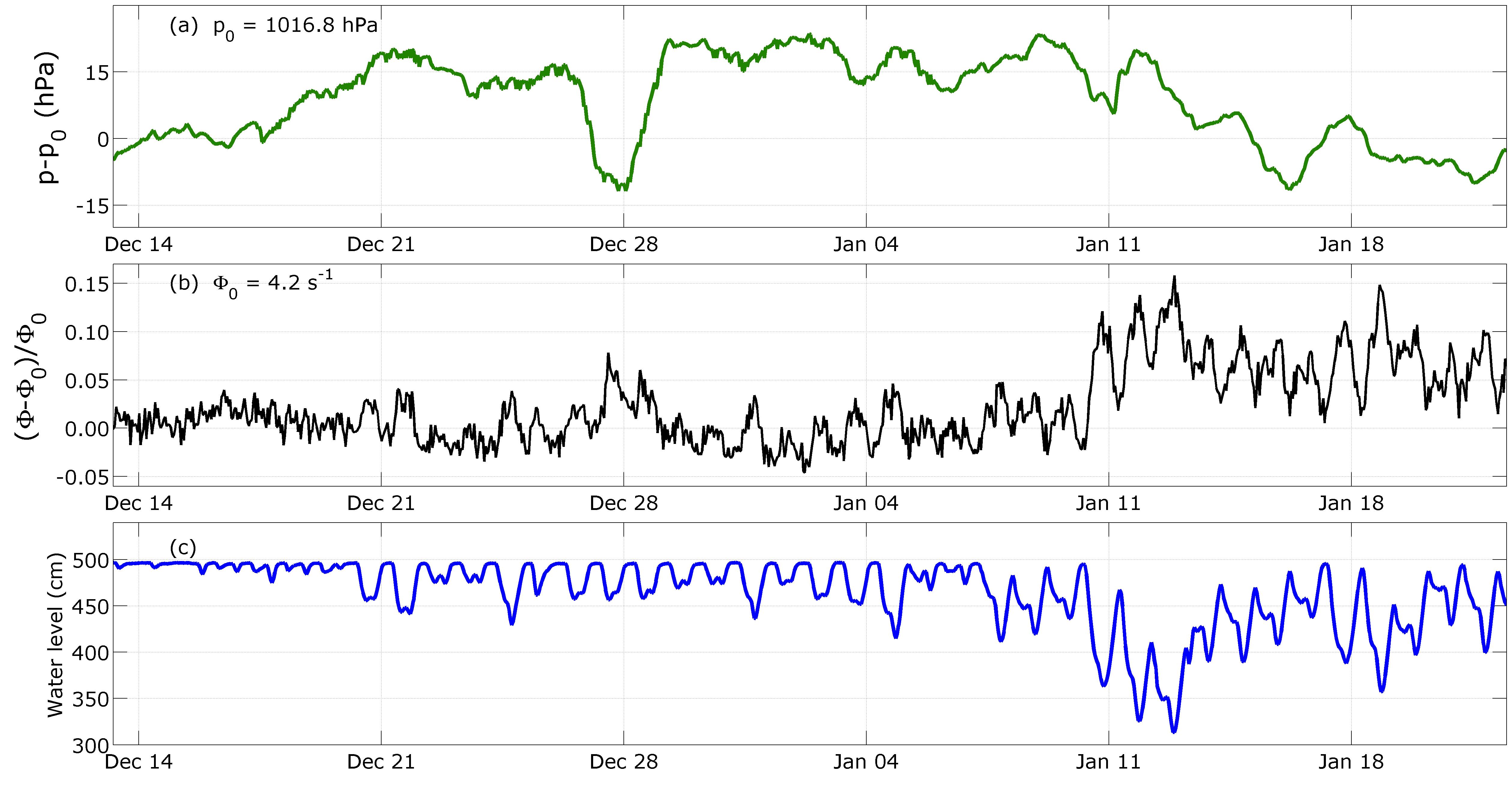}
\caption{
{Hourly averages of the data acquired} during the non-stationary period (December $13^{\mathrm{th}}$ $2014$ to January $22^{\mathrm{nd}}$ $2015$): \textbf{(a)} {Atmospheric pressure time-variations relatively} to $p_0 = 1016.8 \; \mathrm{hPa}$. \textbf{(b)} {Muon flux relative raw time-variations} (black curve), {and corrected} from the atmospheric pressure influence (red curve). \textbf{(c)} {Water level variations}.
}
\label{dataGeneralPeriod}
\end{figure} 

\begin{figure}[tb] 
\centering
\includegraphics[width=0.6\linewidth]{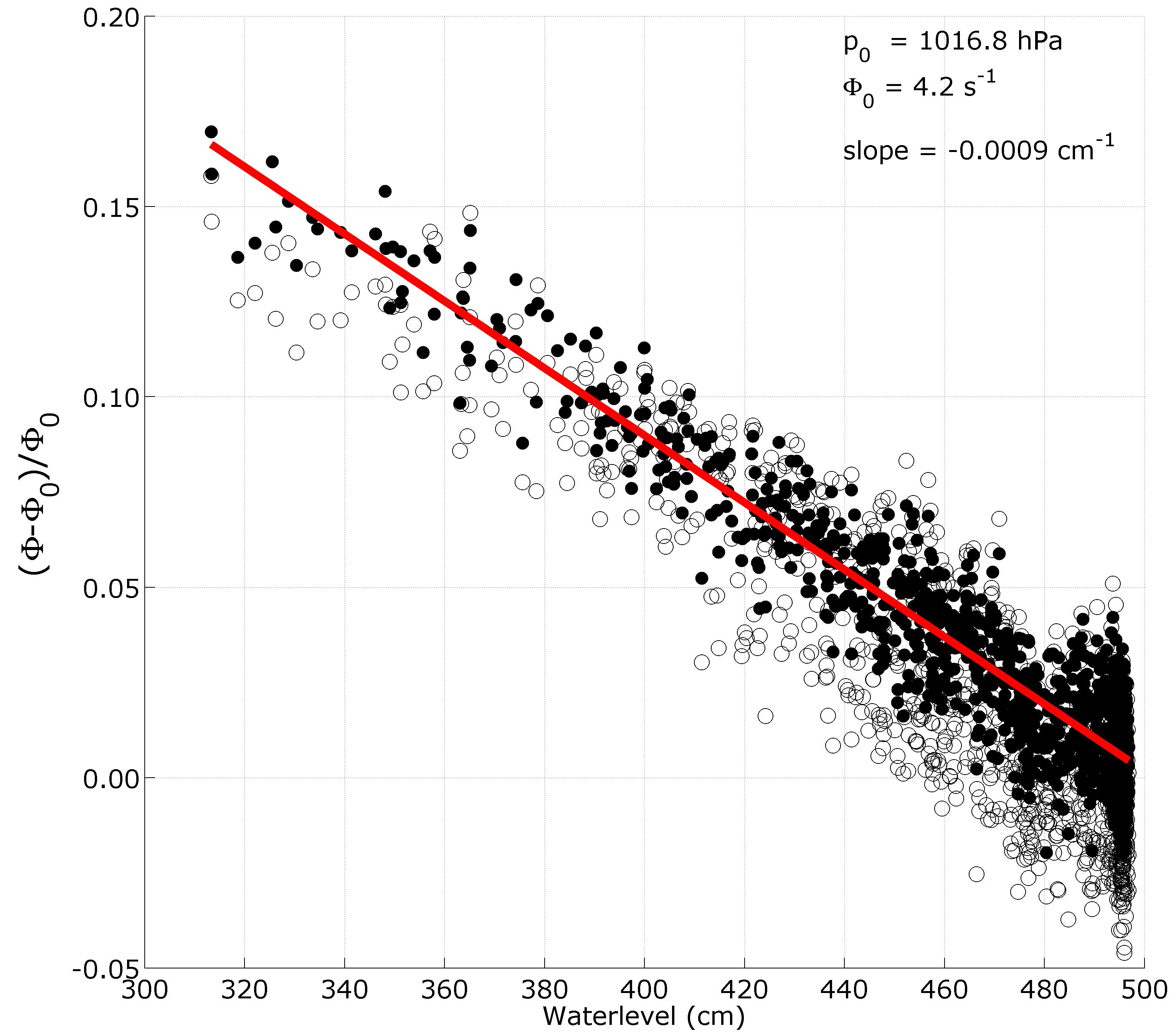}
\caption{
Normalized and centered muon flux as a function of water level. The open circles correspond to the measured muon flux (i.e. black curve in Figure \ref{dataGeneralPeriod}b) and the black dots correspond to the muon flux corrected from the atmospheric pressure influence (equation \ref{pressureCorrection}). {The red straight line is a linear fit to the pressure-corrected points.}
}
\label{fluxWaterLevel}
\end{figure} 

We now consider the data {from the second measurement period, where we observe large water level variations} (Figure~\ref{dataGeneralPeriod}). {It begins} on December $13^{\mathrm{th}}$ $2014$ and ends on January $22^{\mathrm{nd}}$ $2015$. The largest decrease in the water level is up to nearly $200 \hspace{1mm} \mathrm{cm}$ with respect to $h_0$ (Figure~\ref{dataGeneralPeriod}c). During the same period, the {muon flux variations} appear clearly anti-correlated with the water level (Figure~\ref{dataGeneralPeriod}b), and the highest relative flux deviation reaches $15\%$ when the water level is minimum ($\approx 320 \hspace{1mm} \mathrm{cm}$). Meanwhile, the atmospheric pressure variations (Figure~\ref{dataGeneralPeriod}a) also produce conspicuous effects on the muon flux like, for instance, the flux bump that occurs around December $28^{\mathrm{th}}$ during a low-pressure event.

The circles in Figure \ref{fluxWaterLevel} represent the muon flux data versus the water level. Applying the atmospheric correction (\ref{pressureCorrection}) to the muon flux reduces the {data points scattering} and enhances the correlation between the flux and the water level (black dots in Figure \ref{fluxWaterLevel}). {The uncorrected points standard deviation $\sigma = 0.019$ is reduced to $\sigma = 0.011$ for the pressure-corrected data. The uncorrected points standard deviation regularly increases from $\sigma = 0.011$ to $\sigma = 0.021$ when the water level increases from $3$ to $5$ meters while the pressure-corrected points standard deviation remains constant in the whole range of water levels. We explain this feature by the fact that high water levels are more frequent than low levels. Consequently the atmospheric pressure fluctuates in a wider range during the time period of high water level, causing larger muon flux variations.}

A linear fit to the pressure-corrected points is displayed by the red line in Figure \ref{fluxWaterLevel} with a negative slope $\beta_h = -0.0009 \; (0.00002) \; \mathrm{cm}^{-1}$ and an intercept $\Delta\Phi_w = 0.444 \; (0.007)$. The standard deviations assigned to the water level and muon flux are respectively $\sigma_h = 1.7 \; \mathrm{cm}$ and $\sigma_\Phi = 0.0082$. The residuals standard deviation is not reduced when fitting a second-order polynomial, and we adopt a linear relationship to represent the {water level influence} on the muon flux,
\begin{equation}
\frac{\Phi - \Phi_0}{\Phi_0} = \beta_h \times h + \Delta\Phi_w.
\label{waterCorrection}
\end{equation}

\section*{Discussion of SHADOW data analysis}

The data analyzed in the previous Section show that linear relationships (equations \ref{pressureCorrection} and \ref{waterCorrection}) may safely be used to represent the {relative muon flux dependence} with respect to {the atmospheric pressure (Figure \ref{fluxPressure}) and water level (Figure \ref{fluxWaterLevel}) variations}. Owing to the fact that the {opacity fluctuations} produced by $\Delta p = 1 \; \mathrm{hPa}$ and $\Delta h = 1 \; \mathrm{cm}$ are identical (i.e. they represent the same mass of matter), it may be deduced that the $\beta$ coefficients in equations (\ref{pressureCorrection}) and (\ref{waterCorrection}) should be the same. This hypothesis is not supported by our experimental results which indicate that $\beta_p$ is significantly larger than $\beta_h$.

{We explain} the discrepancy between the experimental values for $\beta_p$ and $\beta_h$ by the fact that the atmosphere is not only, like water in the tank, a screen of matter for the muon flux but it is also the place where muons originate \cite{gaisser1990cosmic, grieder2001cosmic, anchordoqui2004high}. Consequently, the {muon flux} at ground level {depends} on both the pressure and the temperature profiles in the atmosphere. For instance, if the atmosphere is warmer, the muon production altitude is higher (roughly at the isobaric level $p = 100 \; \mathrm{hPa}$) and the {muons transit times increase}. Then, muons are more likely to decay before reaching the ground {and thus the relative muon flux decreases} \cite{gaisser1990cosmic, anchordoqui2004high}. This is the so-called negative temperature effect. However, an increase in temperature at the production level decreases the air density, thus reducing the likelihood of pion interactions before their decay into muons. Muon production then increases, and this phenomena is known as the positive temperature effect. Both the pressure and temperature effect upon the flux of muons at ground level may be summarized by \cite{blackett1938instability, duperier1944new},
\begin{equation}
\frac{\Phi - \Phi_0}{\Phi_0} = \beta_p^* \times (p-p_0) + \beta_T^* \times (T-T_0),
\label{atm_fluctuations_1}
\end{equation}
where $T$ is the temperature at the production level and $\beta_p^*$ and $\beta_T^*$ are adjustable coefficients for the pressure and temperature effects respectively. The coefficient $\beta_p^*$ is always negative while $\beta_T^*$ may be either positive or negative depending on the prevailing temperature effect. For the soft muon component ($\leq 10 \; \mathrm{GeV}$) which composes the main part of the particles detected by our telescope in this experimental context, the negative temperature effect dominates and $\beta_T^*$ is expected to be negative. The correlation analysis recently performed by Zazyan et al. \cite{zazyan2015atmospheric} shows that the pressure and temperature effects are positively correlated. Consequently, for the {present measurement conditions}, both $\beta_p^*$ and $\beta_T^*$ are negative and time-correlated. When considering the atmospheric effect alone like in equation (\ref{pressureCorrection}), the $\beta_p$ coefficient actually accounts for both the pressure and temperature effects. This explains why the experimental value found for $\beta_p$ (\ref{pressureCorrection}) is larger than the value of $\beta_h$ (\ref{waterCorrection}).

\section*{Statistical feasibility and limits of opacity monitoring}

\begin{figure} 
\centering
\hspace{-0cm}
\includegraphics[width=0.7\linewidth]{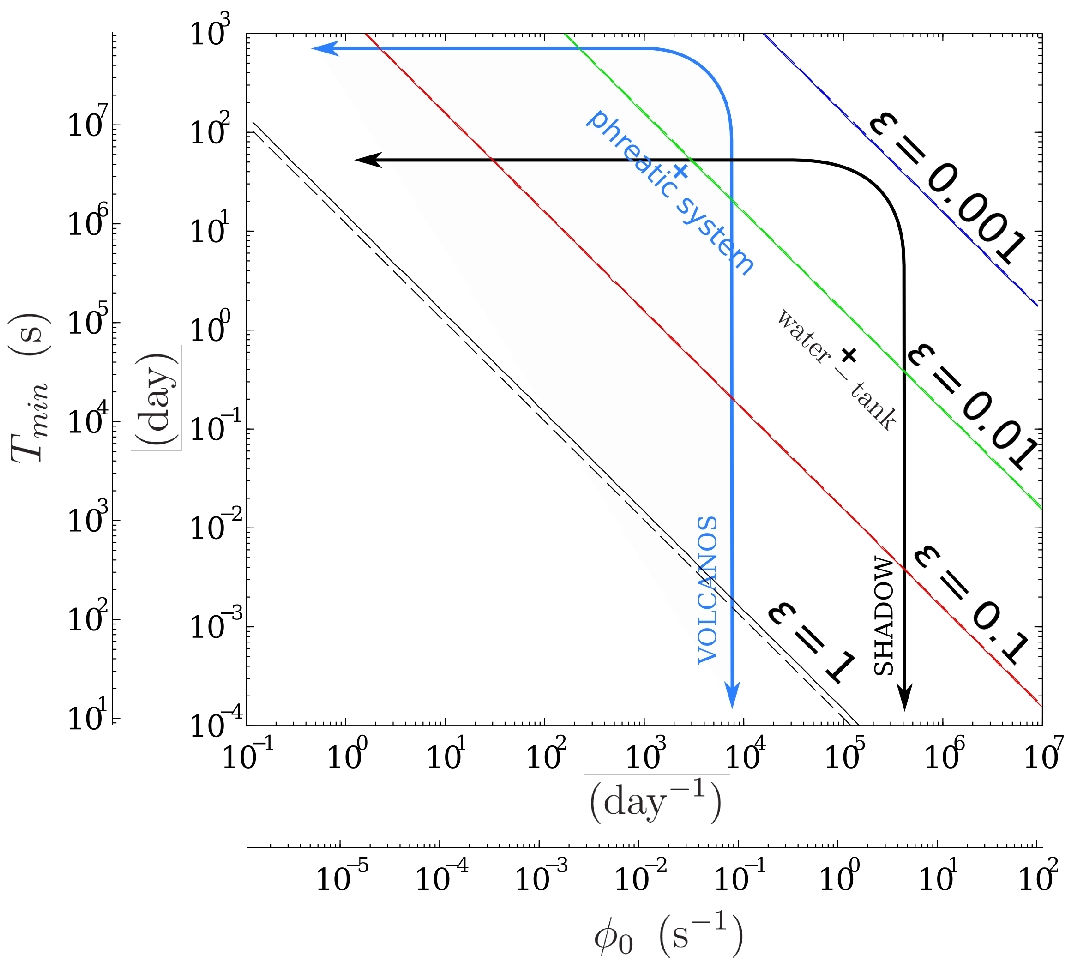}
\caption{
Minimum acquisition time $T_{min}$ versus the average measured flux $\phi_0$ necessary to detect a {$\varepsilon$ flux variation} with $\alpha = 0.05$ (with a $95\%$ confidence level). The straight and dotted lines are the iso-$\alpha$ curves respectively computed with  equation~(\ref{Skellam_1}) and the approximation (\ref{gaussian_2}) . The curved arrows delimit the resolution domains for the SHADOW experiment and typical volcano applications. The horizontal limit marked by the arrows is the measurement whole duration and the vertical limit is the maximum flux measured. The crosses represent likely sources of {muon flux variations}, their coordinates depend on the flux fluctuations amplitude $\alpha$ and their typical period $T_{min}$.
}
\label{feasability_flux_figure}
\end{figure} 

We now address some statistical issues concerning the monitoring of opacity variations like those produced by water-level variations measured during the SHADOW experiment.

Let us assume that $N = N_1 + N_2$ particles are detected by the telescope during a time period $T$, and where $N_1$ and $N_2$ are the number of particles respectively counted during the first and second half of $T$. We want to determine under which conditions $N_1$ and $N_2$ may be considered different at the confidence level $\alpha$. The particle flux difference $\Delta N = N_2-N_1$ obeys a Skellam distribution defined as the difference between two Poisson processes with means $\mu_1$ and $\mu_2$ \cite{skellam1945frequency},
\begin{equation}
\mathcal{S}\left(\Delta N, \mu_1, \mu_2 \right) = \mathrm{e}^{-(\mu_1+\mu_2)}\left(\frac{\mu_1}{\mu_2}\right)^{k/2}\mathrm{I}_{\Delta N}(2\sqrt{\mu_1 \mu_2}),
\label{skellam}
\end{equation}
where $\mathrm{I}_{\Delta N}$ is the modified Bessel function of the first kind.

In the case where $N_2 > N_1$, the hypothesis $\Delta N \neq 0$ may be considered true at the confidence level $\alpha$ if
\begin{align}
\sum_{i = -\infty}^{-1} &  \mathcal{S}\left( i, N_1, N_2 \right) + \frac{1}{2} \times \mathcal{S}\left( 0, N_1, N_2 \right) \le 1 - \alpha \label{Skellam_1}\\
N_1 & = T/2 \times \phi_1 = T/2 \times \phi_0 \times \left(1-\varepsilon/2\right) \label{Skellam_2} \\
N_2 & = T/2 \times \phi_2 = T/2 \times \phi_0 \times \left(1+\varepsilon/2\right). \label{Skellam_3}
\end{align}
{with $\varepsilon$ the flux variation percentage.}

When the inequality (\ref{Skellam_1}) becomes an equality we get $T = T_{min}$, the minimum acquisition time necessary to resolve a flux difference given by the following set of parameters $(\phi_0, \varepsilon, \alpha)$. When $\varepsilon$ is fixed, $T_{min}$ is the best time resolution achievable to observe temporal {relative flux variations} larger than $\varepsilon$. When $T_{min}$ is fixed, we derive the best relative flux variation, $\varepsilon$, detectable on a time-scale larger than $T_{min}$.

Note that if $(N_1,N_2) \gtrsim 10$ the Poisson laws can be approximated with Gaussians and equation~(\ref{Skellam_1}) is simplified to,
\begin{align}
(N_2 & -N_1) - \tilde{\alpha} \times \sqrt{\frac{N_1 \times N_2}{N_1 + N_2}} \ge 0 \label{gaussian_1} \\
& T \ge T_{min} = \frac{\tilde{\alpha}^2 \left( 1 - \varepsilon^2 / 4 \right)}{\varepsilon^2 \times \phi_0} \label{gaussian_2} 
\end{align}
where $\alpha = \mathrm{erf}(\tilde{\alpha})$.

We numerically compute $T_{min}$ from equation~(\ref{Skellam_1}) with a confidence level $\alpha = 0.05$ and represent it on Figure~\ref{feasability_flux_figure} for a range of measured muon flux and variation threshold $\varepsilon = 1, 0.1, 0.01$, and $0.001$ (i.e. $100\%, 10\%, 1\%$, and $0.1\%$). Observe that the approximation (\ref{gaussian_2}) is suitable for our range of applications, $T_{min}$ will be underestimated starting from $\varepsilon \gtrsim 0.5$ which implies $N \approx 20$.

Figure \ref{feasability_flux_figure} shows that to detect a daily variation in the {muons count} of $2\%$ ($\varepsilon = 0.02$), as is typically observed in the SHADOW experiment, an average flux $\phi_0 > 2  \; \mathrm{s^{-1}}$ must be measured. This solution is represented by the black cross labelled ``water tank'' on Figure~\ref{feasability_flux_figure}. The lower-left domain delimited by the curved black arrow in Figure~\ref{feasability_flux_figure} represents the solution-domain for time scales and opacity variations of the SHADOW experiment category. This solution-domain is the region where flux variations can be resolved at a high confidence level. The {arrow horizontal branch} is limited by the {experiment duration}, and the vertical branch is placed at a level corresponding to the maximum flux that can be measured by the telescope. This latter quantity {increases with the telescope acceptance, e.g.  by increasing the angular aperture (i.e. by reducing the distance between the detection matrices), or by grouping several lines of sight, or by using instruments with a larger detection surface}.

The feasibility domain for a typical volcano experiment is also represented on Figure~\ref{feasability_flux_figure} and delimited by the blue curved arrow. Note that for this kind of experiments we have a longer acquisition time and a tiny measured flux as the total opacity of the geological body facing the telescope is much bigger than for the SHADOW experiment: about $1000 \; \mathrm{m.w.e.}$ for a volcanic lava dome versus $5 \; \mathrm{m.w.e.}$ for the water tank.

We can rewrite equation~(\ref{Skellam_1}) into a form more suitable for radiography applications by replacing the flux fluctuations by opacity fluctuations,
\begin{align}
N &= T \times \phi \left( \mathcal{T} , \varrho_0 , \theta \right) \label{tomo_1} \\
N_1 &= T/2 \times \phi \left( \mathcal{T}, \varrho_0 \times (1-\varepsilon /2), \theta \right) \label{tomo_2} \\
N_2 &= T/2 \times \phi \left( \mathcal{T}, \varrho_0 \times (1+\varepsilon /2), \theta \right) \label{tomo_3},
\end{align}
where the muon flux $\phi$, is explicitly written to depend on telescope acceptance $\mathcal{T}$, opacity $\varrho_0$, and zenith angle $\theta$. As before, $\varepsilon$ represents the variation of opacity relative to the average opacity $\varrho_0$. We warn the reader that a given $\varepsilon$-variation of $\varrho$ corresponds to a much larger $\varepsilon$-variation of $\phi$. Putting equations~(\ref{tomo_2}) and (\ref{tomo_3}) in equation (\ref{gaussian_1}) we obtain the following feasibility condition,
\begin{equation}
T \ge T_{min}(\varrho_0,\varepsilon,\theta,\mathcal{T}) = \frac{2 \times \tilde{\alpha}^2 \times \phi_1 \times \phi_2}{(\phi_2 - \phi_1)^2 \times (\phi_2 + \phi_1)} \label{feasability_opacity}
\end{equation}
where $T_{min}$ is the {measurement period minimum duration} necessary to resolve the sought opacity variation. Note that the feasibility formula from Lesparre et al. \cite{lesparre2010geophysical} is the first order development of equation (\ref{feasability_opacity}).

\begin{figure} 
\centering
\includegraphics[width=0.7\linewidth]{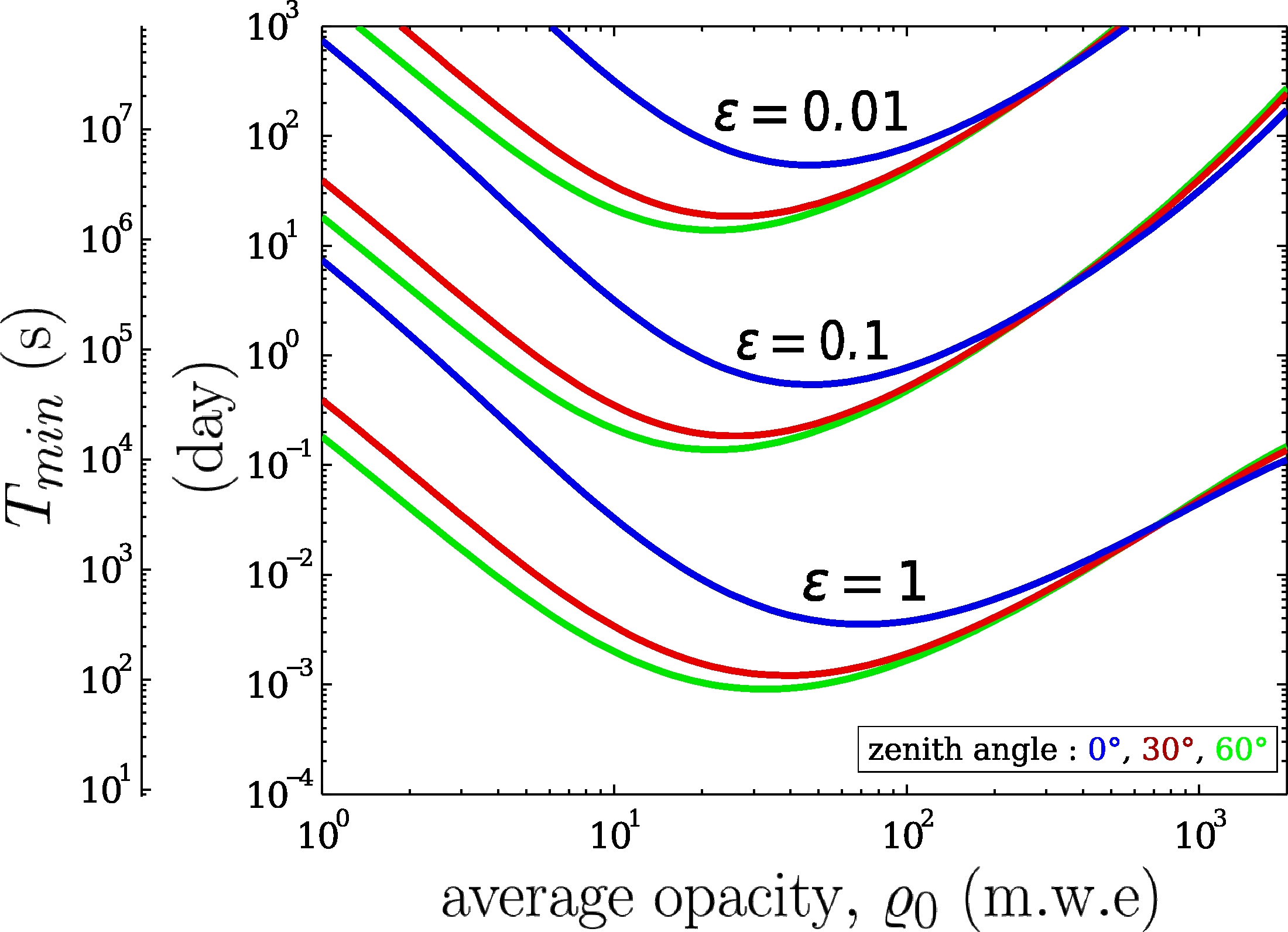}
\caption{
Minimum acquisition time $T_{min}$ as a function of the average opacity $\varrho_0$ to detect an $\varepsilon$ fluctuation at the $\alpha = 0.05$ confidence level. The three curves (resp. blue, red, green) correspond to three different observation zenith angles (resp. $0^{\circ}$, $30^{\circ}$, $60^{\circ}$) and are computed for $\mathcal{T} = 10 \; \mathrm{cm^2.sr}$ using the modified Gaisser model from Tang et al (2006).
}
\label{feasibility_opacity_figure}
\end{figure}

A subset of $T_{min}$ solutions of equation (\ref{feasability_opacity}) is represented on Figure \ref{feasibility_opacity_figure} for the confidence level $\alpha = 0.05$, for zenith angles $\theta = 0^{\circ}$, $30^{\circ}$, and $60^{\circ}$, and opacity variations $\varepsilon = 100\%$, $10\%$, $1\%$. An acceptance, $\mathcal{T} = 10 \; \mathrm{cm^2.sr}$, typical of our telescopes has been used in the computation. Roughly, a one order of magnitude {$\varepsilon$ variation} induces a {$T_{min}$ change} by two orders of magnitude. 

Observe that there is an optimal opacity range where the measurement time, i.e. the time resolution that is achievable, is minimum to resolve a given opacity variation. The optimal opacity range depends on the zenith angle and goes roughly from $40$ to $100 \; \mathrm{m.w.e}$ for $\theta = 0^{\circ}$, and from $20$ to $40 \; \mathrm{m.w.e}$ for $\theta = 60^{\circ}$. For low-opacity conditions, measurements at high zenith angles are wise to optimize the time resolution. This is particularly conspicuous for the SHADOW experiment where the average opacity $\varrho_0 \approx 5 \; \mathrm{m.w.e}$ and $\varepsilon \approx 10\%$. For these parameters, Figure \ref{feasibility_opacity_figure} gives $T_{min} > 1 \; \mathrm{day}$ is necessary at $\theta = 0^{\circ}$ to resolve the fluctuations while $T_{min} > 0.2 \; \mathrm{day}$ is sufficient at $\theta = 60^{\circ}$. The {time resolution strong dependence} with respect to the zenith angle disappears at larger opacities $\varrho_0 > 500 \; \mathrm{m.w.e}$ like those encountered in volcano muon radiography.

\section*{Discussion}

Muon radiography is a powerful method to monitor opacity/density variations inside geological bodies. Noticeable advantages of the method are the possibility to remotely radiography unapproachable dangerous volcanoes and to image the density distribution of large volumes from a single view-point \cite{nagamine1995method, lesparre2012density, jourde2013experimental}. Muon radiography is entering an era of precision measurements not only for structural imaging but also for dynamical monitoring purposes. Some monitoring experiments have been performed on active volcanoes that demonstrate the usefulness of such measurements to constrain the evolution of eruption crisis \cite{tanaka2014radiographic}. However, as shown above, monitoring opacity variations is subject {to external sources of bias, and statistical and experimental constraints} that limit the achievable resolution. Understanding these limits is of primary importance to improve the method and to assess the {muon radiography monitoring feasibility and validity}.

Experimental constraints are partly dictated by statistical considerations, and mainly come from the telescope acceptance that limits the maximum flux which fixes the {resolution domain right boundary} in Figure \ref{feasability_flux_figure}. This boundary may be moved rightward by increasing the acceptance $\mathcal{T}$ of the instrument. Recalling that $\mathcal{T}$ is expressed in $[\mathrm{cm^2sr}]$, the acceptance may be augmented by several means: 1) increasing the solid angle encompassed by the instrument by reducing the distance between the detection matrices; 2) increasing the detection surface by coupling several telescopes (actually our telescopes may be merged into a single one); 3) grouping lines of sight to increase both the detection surface and the solid angle at the price of reducing the angular resolution of the radiographies. In the present study, the latter solution {was} retained and all lines of sight {were} merged to obtain an effective acceptance of $630 \; \mathrm{cm^2sr}$.

Statistical constraints bound the resolution domain of a given experiment (Figure \ref{feasability_flux_figure}), and the main concern when doing measurements is to ensure that the monitored phenomena fall inside the boundaries. As will be discussed in the next paragraph, the telescope configuration may be adapted to comply with the {ongoing experiment objectives}. As shown in the preceding sections, the statistical constraints are quite different whether the opacity is high or low. This is conspicuous in Figure (\ref{feasibility_opacity_figure}) where the feasibility solutions for $T_{min}$ strongly differ in the low- and high-opacity domains. It is remarkable that high-opacities variations are equally resolved whatever the zenith angle while, instead, the resolution for low-opacities strongly depends on this angle. Another conspicuous feature present in Figure (\ref{feasibility_opacity_figure}) is the existence of an optimal medium-opacity range $\varrho \approx 50 \pm 30 \; \mathrm{m.w.e.}$ where telescopes offer their best performance. {These two effects are due to} the {cosmic muon energy spectrum nature} \cite{gaisser1990cosmic,hebbeker2002compilation,tang2006muon}, and changing the {telescope acceptance} has no effect on the optimal opacity values but only changes $T_{min}$ by translating the solution curves of Figure (\ref{feasibility_opacity_figure}) either upward (decrease of acceptance) or downward (increase of acceptance).

\section*{Methods}

\begin{figure}[bt] 
\centering
\includegraphics[width=0.5\linewidth,angle=-90]{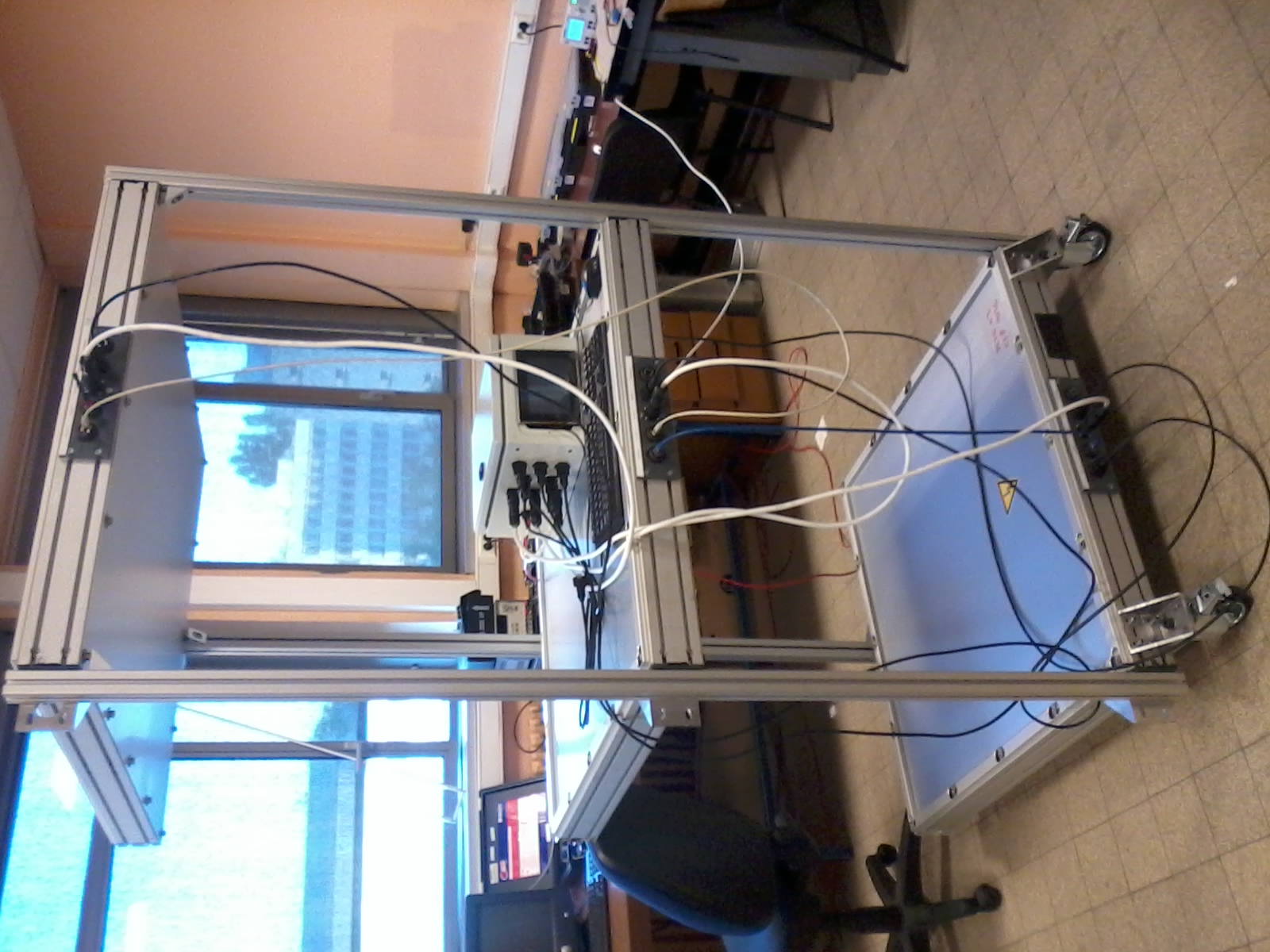}
\caption{
Picture of the muon telescope used for the SHADOW experiment, here during the open-sky calibration phase. The three detection matrices are horizontal. The calibration gives access to the effective acceptance. The control box embedding a mini-PC, a common clock distribution system, a network switch is visible on the middle matrix.
}
\label{telescope_picture}
\end{figure} 

\begin{figure}[bt] 
\centering
\includegraphics[width=0.5\linewidth]{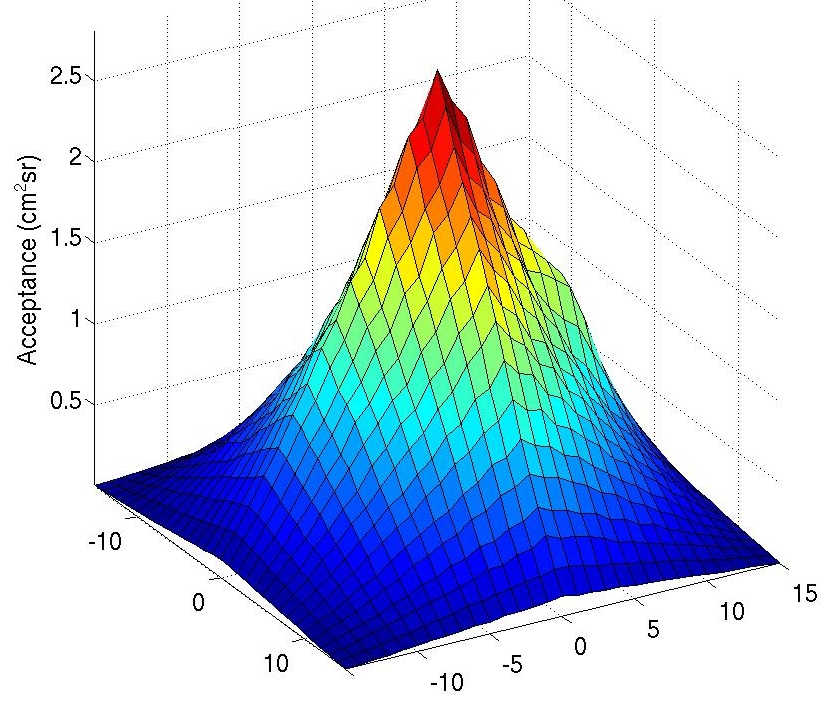}
\caption{
{Telescope experimental acceptance} for the configuration shown in Figure (\ref{experimental_setup_figure}). The {acceptance maximum value}, $\mathcal{T}_{\max} = 2.80 \; \mathrm{cm}^2\mathrm{sr}$ is obtained for the line of sight perpendicular to the detector planes and corresponding to $(x,y) = (0,0)$. The $x$ and $y$ coordinates represent the horizontal offsets between the pixels defining a given line of sight of the telescope (one pixel in the upper detection matrix, and the other one in the lower matrix). The acceptance integrated over the {instrument entire detection surface} equals $\mathcal{T}_{\mathrm{int}} = 630 \hspace{1mm} \mathrm{cm}^2\mathrm{sr}$ for a solid angle aperture $\Omega_{\mathrm{int}} = 0.161 \hspace{1mm} \mathrm{sr}$.
}
\label{telescopeAcceptance}
\end{figure} 

The muon count series analysed in the present study were acquired with one of our standard telescopes shown in Figure \ref{telescope_picture} \cite{lesparre2012design,jourde2013experimental,marteau2014implementation}. The picture was taken  during an open-sky calibration phase where the muon count serves to determine the efficiency of the scintillator bars forming the detection matrices. Each matrix is formed by an assemblage of two sets of $16$ bars arranged perpendicularly {to} obtain {a $16 \times 16$ square $5 \times 5 \; \mathrm{cm}^2$ pixels array}. The {telescope upper and lower matrices} allow $31 \times 31$ {pixels combinations}, i.e. $961$ distinct lines of sight. The distance between the matrices may be changed to adapt the solid angle spanned by the trajectories. In the present study, the distance was tuned to encompass the entire water tank (Figure \ref{experimental_setup_figure}).

Once geometrically configured, the telescope is totally characterised by its acceptance function $\mathcal{T}_i \; [\mathrm{cm}^2\mathrm{sr}]$ which relates the muon count, $N_i$, to the {muon flux}, $\partial\phi \; [\mathrm{s}^{-1}\mathrm{cm}^{-2}\mathrm{sr}^{-1}]$ received by the telescope in its $i^\mathrm{th}$ line of sight,
\begin{align}
N_i & = T \times \int_{4\pi} \mathcal{P}_i (\varphi,\theta) \times \partial\phi (\varrho,\varphi,\theta) \times \mathrm{d} \Omega, \\
  & = T \times \mathcal{T}_i \times \partial\phi_i,
\label{equationMethod1}
\end{align}
where $T$ is the {acquisition duration}, $\mathcal{P}_i \; [\mathrm{cm}^2]$ is the {line of sight detection surface function}, $\mathcal{T}_i$ is the integrated acceptance, and $\partial\phi_i$ is the muon flux in the {line of sight central direction}. It must be understood that $\partial\phi \; [\mathrm{s^{-1}cm^{-2}sr^{-1}}]$ is the differential {muon flux that reaches} the instrument after crossing the target. Consequently, $\partial\phi$ depends both on the open sky differential flux $\partial\phi (\varrho = 0,\varphi,\theta)$ and on the muon absorption law inside matter. These are determined through experiments \cite{sagisaka1986atmospheric, ambrosio1997seasonal, adamson2010observation, tilav2010atmospheric, hebbeker2002compilation, motoki2003precise, poirier2011periodic}, theoretical works \cite{tang2006muon} or thanks to Monte-Carlo simulations \cite{heck1998corsika, wentz2003simulation} depending on the precision expected and the available information.

Figure (\ref{telescopeAcceptance}) shows the {telescope acceptances} $\mathcal{T}_i$ for $i = 1,\cdots,961$ used in the SHADOW experiment. This acceptance function is determined experimentally to account for the {detection matrices deffects}, mainly imperfect optical couplings at {the scintillator bars outputs} and on the {multichannel photomultiplier front}. {The latter causes} the distortions visible in the {Figure (\ref{telescopeAcceptance}) 3D plot}. In practice, the {acceptance computation} is performed by measuring the ``open-sky'' muons flux coming from the zenith.

The {detected particles number} $N$ may be increased by grouping several adjacent lines of sight belonging to a subset $\mathcal{E}$,
\begin{equation}
N_{\mathcal{E}} = T \times \sum_{i \in \mathcal{E}} \mathcal{T}_i \times \partial \phi_i = \sum_{i \in \mathcal{E}} N_i.
\label{N_particles_detected}
\end{equation}
It results in an {acceptance increase} {and thus} a better time resolution. The counterpart is {an angular resolution degradation} induced by the merging of the small solid angles spanned by the trajectories. In the present study, the entire solid angle spanned by the telescope trajectories were grouped to obtain a total acceptance $\mathcal{T}_{\mathrm{total}} = 630 \; \mathrm{cm}^2\mathrm{sr}$. Such a large acceptance dramatically improves the time resolution which falls to the order of tens of minutes in the case of the SHADOW experiment.

\section*{Acknowledgements}
This study is part of the DIAPHANE project ANR-14-CE 04-0001. We acknowledge the financial support from the UnivEarthS Labex program of Sorbonne Paris Cit\'e (\textsc{anr-10-labx-0023} and \textsc{anr-11-idex-0005-02}). We would like to thank also the members of the SIEPC, Tignieu-Jameyzieu, France, for their help, support and the access to the data. In particular we thank Mr Gilbert Pommet and Mr R\'emi Cachet. The results presented in this paper rely on rely on geomagnetic Kp indices calculated and made available by ISGI Collaborating Institutes and on the data collected at Chambon-la-Foret magnetic observatory. We thank the Institut de Physique du Globe de Paris, for supporting its operation and the INTERMAGNET network and ISGI (isgi.unistra.fr). This is IPGP contribution ****.

\section*{Author contributions statement}
K.J., J.M. and D.G. conceived the experiment, all authors designed and constructed the apparatus, J.M. conducted the experiment, K.J., J.M. and D.G. analysed the data. All authors reviewed the manuscript.

\section*{Additional information}

\textbf{The authors declare no competing financial interests.}

The corresponding author is responsible for submitting a \href{http://www.nature.com/srep/policies/index.html#competing}{competing financial interests statement} on behalf of all authors of the paper. This statement must be included in the submitted article file.



\end{document}